%% file: main.tex
\makeatletter
\def\input@path{
{/pic_bib/}
}
\makeatother
\documentclass{WileyMSP-template}
\usepackage{amssymb}
\usepackage{mathrsfs}
\usepackage{float} 
\usepackage{placeins}
\usepackage{bm}
\usepackage{amsfonts}
\usepackage{xcolor}
\usepackage{dirtytalk}
\usepackage{calc}
\usepackage{pifont}
\usepackage{dsfont}
\usepackage{graphicx}
\usepackage{array}
\newcolumntype{C}[1]{>{\centering\arraybackslash}p{#1}}
\usepackage{float}

\input{pic_bib/my_style}
\usepackage{ragged2e}
\justifying

\begin{document}

\pagestyle{fancy}
\rhead{\includegraphics[width=2.5cm]{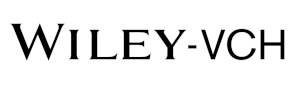}}
\noindent\title{Nonreciprocal electromagnetic wave manipulation via a single reflection}
\maketitle

\noindent\author{Lu Wang*}

\dedication{}

\begin{affiliations}
\noindent Dr. Lu Wang\\
Address: State Key Laboratory of Magnetic Resonance and Atomic and Molecular Physics, Wuhan Institute of Physics and Mathematics, Innovation Academy for Precision Measurement Science and Technology, Chinese Academy of Sciences, Wuhan 430071, China\\
Email Address: lu.wangthz@outlook.com
\end{affiliations}
\noindent\keywords{nonreciprocity, electric field manipulation, photonics, far-field}

\begin{abstract}
Electric field manipulation plays a key role in applications such as electron acceleration, nonlinear light-matter interaction, and radiation engineering. Nonreciprocal materials, such as Weyl semimetals, enable the manipulation of the electric field in a full photonic manner, owing to their intrinsic time-reversal symmetry breaking, leading to asymmetric material response for photons with $+\mathbf{k}$ and $-\mathbf{k}$ momenta.
Here, the results suggest that a simple planar interface between semi-infinite air and a nonreciprocal material can achieve spatio-temporal manipulation of the electric field. In particular, this work presents three compelling scenarios for electromagnetic wave manipulation: radiation pattern redistribution (with closed-form expressions provided), carrier-envelope-phase control, and spatial profile control. The presented results pave the way for electric field manipulation using pattern-free nonreciprocal materials.
\end{abstract}

\section{Introduction} 
Electric field manipulation plays a key role in applications such as electron acceleration \cite{wang2022spatio}, nonlinear light-matter interaction \cite{wang2020tilted}, and radiation engineering \cite{wang2023maximal}. Though with crucial impacts, the manipulation of the electric field is commonly achieved by complicated optical elements such as metamaterial \cite{bai2022radiation}, grating \cite{wu2021strong,shayegan2022nonreciprocal}, and wave plate \cite{deng2022recent}. These complicated structures bring challenges in manufacturing and thus limit related applications. In addition, it is worth noting that the majority of optical elements exhibit limited responsiveness within specific frequency ranges. Consequently, even if these elements are accessible in the optical regime, acquiring them for long wavelengths, proves to be exceedingly difficult due to significant loss during propagation through the optical elements and the pronounced diffraction effects experienced by long wavelengths.

In light of these obstacles, I propose an alternative method that utilizes reflection on planar surfaces to manipulate electric fields. The potential applicability of this approach spans a broad frequency range, encompassing terahertz to optical frequencies. It enables effortless engineering in a purely photonics-based manner, opening up exciting opportunities in this field. This can be realized via nonreciprocal materials such as Weyl semimetals.

To comprehensively understand the spatio-temporal behavior of the electric field when influenced by nonreciprocal materials, I investigate three different scenarios. In the first scenario, I explore how the presence of the nonreciprocal material influences the radiation pattern of an ideal dipole, where closed-form expressions are presented. Furthermore, I examine how the nonreciprocal material affects the carrier-envelope phase and phase-front distortions of the electric field, which are relevant for practical applications. The presented results pave the way for electric field manipulation using pattern-free nonreciprocal materials.

\section{Theoretical model}
To achieve the proposed aim i.e. electromagnetic wave manipulation using a planar structure, the nonreciprocal material is the key. Weyl semimetals, a newly discovered class of quantum materials, would serve the purpose. Weyl semimetals possess unique intrinsic time-reversal symmetry breaking, which leads to nonreciprocity. In the momentum space, Weyl nodes appear in pairs with opposite chirality, where the Weyl points are the sink or source of the Berry curvature \cite{chang1996berry,shen2017topological}. Several Weyl semimetals have been experimentally discovered such as $\text{WP}_\text{2}$, $\text{HgCr}_2\text{Se}_4$, $\text{TaAs}$ and $\text{Co}_3\text{Sn}_2\text{S}_2$  \cite{kumar2017extremely,wan2011topological,xu2011chern,okamura2020giant}, which are ready to be explored and implemented. The material property of the nonreciprocal material can be characterized via the permittivity tensor \cite{kotov2018giant}
\begin{equation}
\bm{\epsilon}=
    \begin{pmatrix}
        \epsilon_{d} & 0 &  \epsilon_{xz}\\
        0 & \epsilon_{d}& 0   \\
        -\epsilon_{xz} & 0 &  \epsilon_{d}
    \end{pmatrix}\label{eq:eps},
\end{equation}
where without loss of generalization, the tensor elements are chosen to be $\epsilon_d=1.5+0.3\,i$ and $\epsilon_{xz}=1.5\,i$, which are standard values. Note that Equation (\ref{eq:eps}) is written in the Cartesian coordinates with azimuthal angle $\phi=0$ (see  \textbf{Figure \ref{fig:dipole}}).

The reflection of the nonreciprocal material is calculated by the transfer-matrix method \cite{wang2023maximal,berreman1972optics,passler2017generalized}. The propagation of the fields is derived using the dyadic Green function. Detailed analyses related to the dyadic Green function can be found in the supplementary material (referred to as SM in the following context).

\section{Results}
\subsection{Radiation distribution manipulation}
\begin{figure}[h]
\centering
  \includegraphics[width=0.4\linewidth]{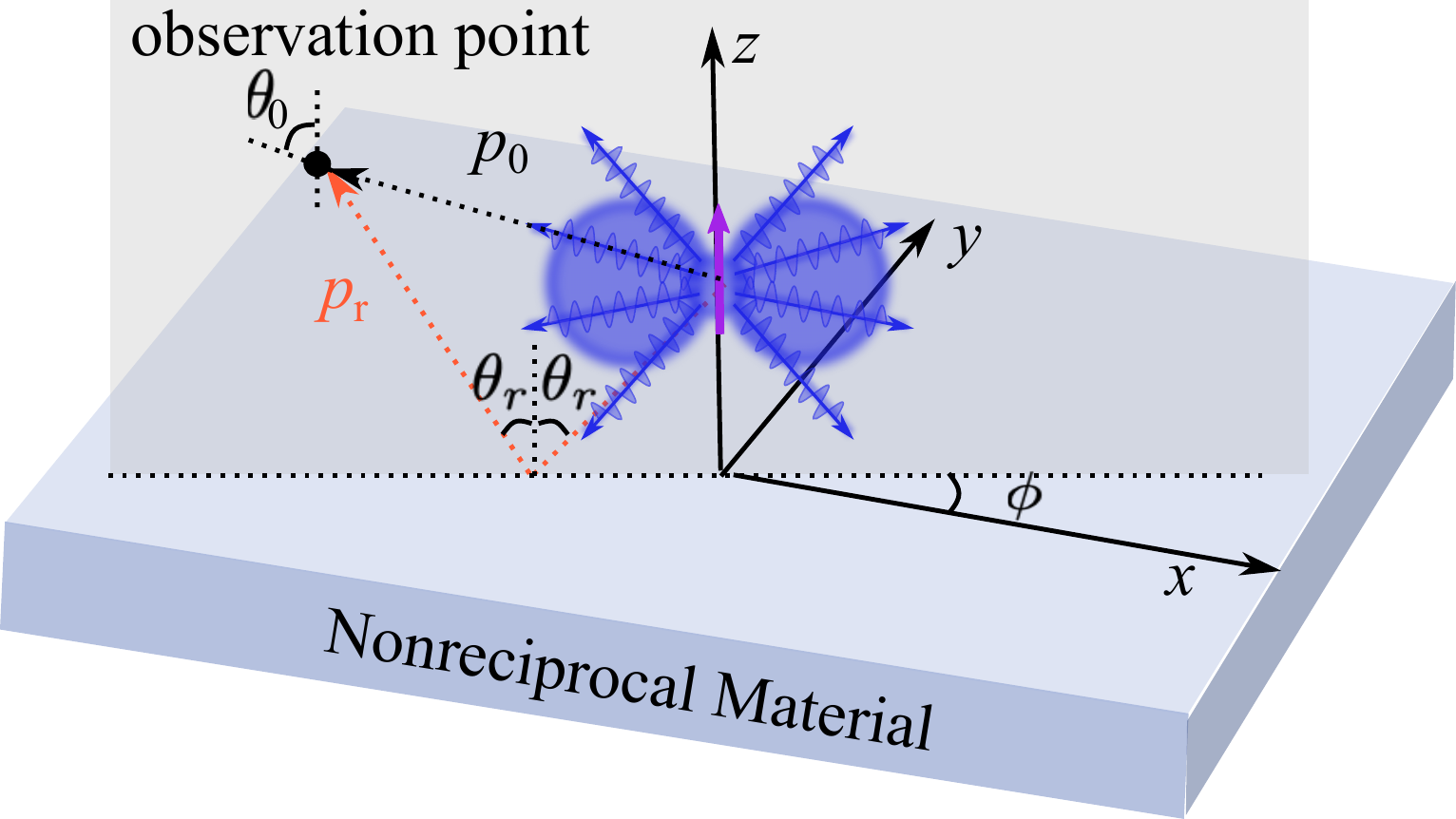}
\caption{Illustration of the dipole emitter $\bm{\mu}$ in the presence of the nonreciprocal material. The dipole is located at $(0,0,z_0)$ and dipole moment is aligned along the $z$ axis $\bm{\mu}=(0,0,\mu_z)$. The observation point is denoted by a black dot $r=(x,y,z)$. In the far-field limit, there are only two values of momenta responding to two paths $p_0$ and $p_r$ that are relevant. The path directly links the source and the observation point is represented by $p_0$. The path that reaches the observation point via reflection off from the nonreciprocal material surface is represented by $p_r$. The variables $\theta_0$ and $\theta_r$ are polar angles of $p_0$ and $p_r$, respectively. }\label{fig:dipole}
\end{figure}
I focus on the ideal dipole as the radiation source since in principle, the results can be extended to the emission of arbitrary objects using the discrete dipole approximation \cite{ekeroth2017thermal}. The propagation of the radiation is governed by the dyadic Green function (see SM Sections S2 and S3). 

With the ideal dipole (point source) located at $r_0=(0,0,z_0)$, and the dipole moment $\bm{\mu}=(0,0,\mu_z)$, one can write the total electric field as
 \begin{equation}\label{eq:E_Green}
\mathbf{E}=\omega_0^2\mu_0\mathbf{G}^E\bm{\mu}=\omega_0^2\mu_0\left(\mathbf{G}_0^E+\mathbf{G}_r^E\right)\bm{\mu}=\mathbf{E}_0+\mathbf{E}_r,
 \end{equation}
where $\mathbf{G}^E$ is a rank two tensor and represents the dyadic Green function, $\omega_0=2\pi c/\lambda_0$ is the angular frequency, $c$ is the speed of light in vacuum, $\lambda_0$ is the wavelength of the radiation in the vacuum, and $\mu_0$ is the vacuum permeability. The subscripts "0" and "r" represent the initial electric field radiated by the dipole and field after the interaction (reflection) with the nonreciprocal material, respectively.

Commonly the Weyl identity is used to describe the dyadic Green function, which requires integrating over the entire $k_x,\, k_y$ momentum space \cite{zhang2020radiative,hu2023high}. This double integral can be cumbersome to analyze the underlying physics and needs high computational demand for implementations. Here, with the far-field limit and the stationary phase approximation \cite{mandel1995optical}, I present closed-form expressions (SM Section S3.3 ). 
\begin{figure*}[h]
\centering
  \includegraphics[width=1\linewidth]{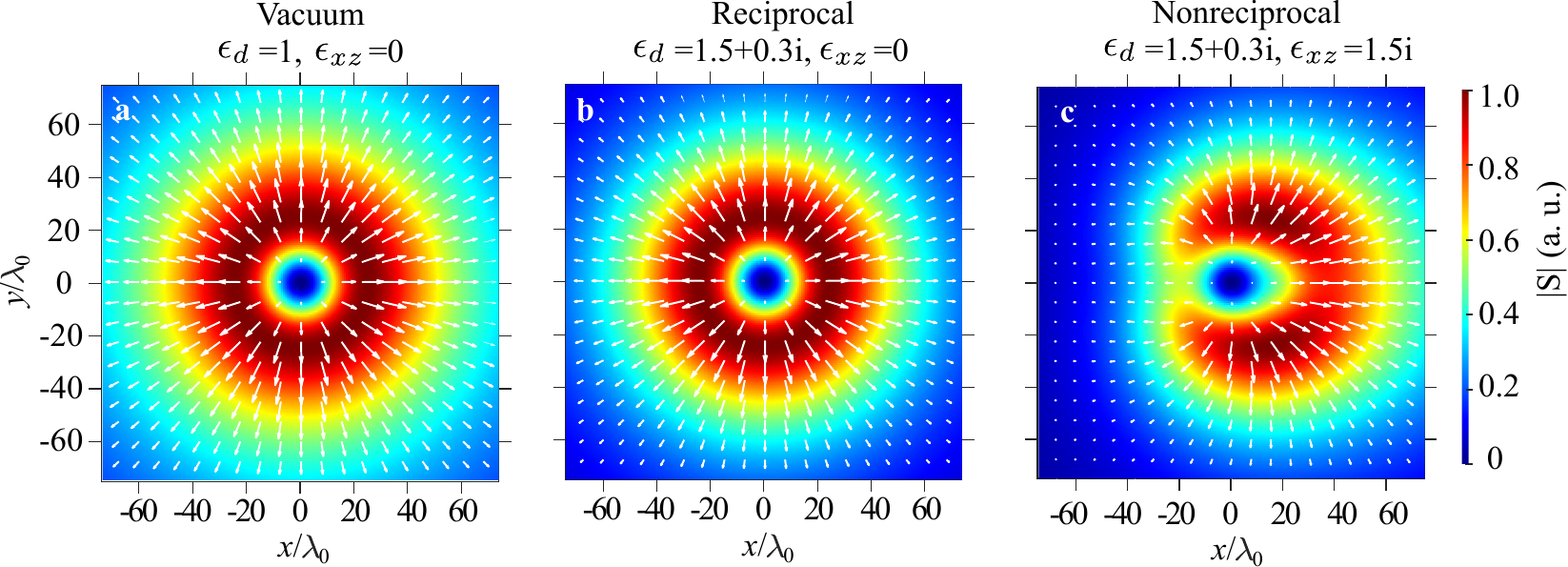}
\caption{Radiation patterns of the ideal dipole in the presence of vacuum (\textbf{a}), reciprocal material (\textbf{b}), and nonreciprocal material (\textbf{c}). The observation plane is chosen to be $z=0.1\lambda_0$. The dipole location $z_0=25\lambda_0$. The amplitude of the time-averaged Poynting vector is represented by $|{S}|$, where $\bm{S}=(S_x,S_y,S_z)=\frac{1}{2}\text{Re}[\bm{E}\times \bm{H}^*]$. The white arrows on each panel represent vector $(S_x, S_y)$. }\label{fig2}
\end{figure*}
The results obtained indicate that the final radiation pattern can be attributed to two distinct propagation paths, denoted as $p_0$ and $p_r$ (see {Figure \ref{fig:dipole}}). The $p_0$ path corresponds to $\mathbf{G}_0^E$, where the radiation travels directly from the dipole to the observation point, covering a distance denoted as $r_0=\sqrt{x^2+y^2+(z-z_0)^2}$. On the other hand, the $p_r$ path corresponds to $\mathbf{G}_r^E$, where the radiation from the dipole reaches the surface of the nonreciprocal material and is subsequently reflected towards the observation point. The distance covered along this path is denoted as $r_r=\sqrt{x^2+y^2+(z+z_0)^2}$. To facilitate further analysis, one can define $f_n={\omega_0^2\mu_0\mu_z\sin{(\theta_n)}\exp{(ik_0r_n)}}/{(4\pi r_n)}$ with $n\in\{0,r\}$, where "0" and "r" again represent $p_0$ and $p_r$, respectively. The analytical expressions of the electric fields can be written as: 
\begin{align}
 \mathbf{E}_0=&f_0\left\{-\cos{(\theta_0)}\left[\cos{(\phi)}\hat{x} +\sin{(\phi)}\hat{y}\right]+\sin{(\theta_0)}\hat{z}\right\} =-f_0\hat{\theta}_0
 \label{eq:E0}\\
\mathbf{E}_r=&f_r\left\{ 
 \left[-r_{pp}\cos{(\theta_r)}\cos{(\phi)}+r_{ps}\sin{(\phi)}\right]\hat{x} -\left[r_{pp}\cos{(\theta_r)}\sin{(\phi)}+r_{ps}\cos{(\phi)}\right]\hat{y}+r_{pp}\sin{(\theta_r)}\hat{z}\right\}  \nonumber\\
  =&-f_r\left( r_{pp}\hat{\theta}_r+r_{ps}\hat{\phi} \right). \label{eq:Er}
\end{align}
The unit vector in the Cartesian coordinates is written as $(\hat{x},\hat{y},\hat{z})$ and the unit vector in the Spherical coordinates is written as $(\hat{r},\hat{\theta}_n,\hat{\phi})$, $n\in\{0,\text{r}\}$. The relation between the unit vectors in Cartesian and spherical coordinates can be found in SM Equation S26-S30.  In this context, $\theta_0$ corresponds to the polar angle associated with $p_0$, whereas $\theta_r$ represents the polar angle of $p_r$ and also serves as the angle of incidence at the surface of the nonreciprocal material. Additionally, $\phi$ denotes the azimuthal angle. The reflection coefficient is denoted by $r_{ij}$ with $i,\,j\,\in\{s,p\}$ and the $r_{sp}$, for example, represents the reflection from s- to p-polarized light.

Since the dipole radiation in the far-field only contains p-polarisation \cite{novotny2012principles_dipole}, $r_{ss}$ and $r_{sp}$ do not show up in Equation (\ref{eq:Er}). A more general expression can be seen in SM Section 3. The results for radiation patterns in the presence of different materials are shown in \textbf{Figure \ref{fig2}}. The radiation pattern is defined as the amplitude of the time-averaged Poynting vector $|\bm{S}|$, where $\bm{S}=\frac{1}{2}\text{Re}[\bm{E}\times \bm{H}^*]$, $\bm{H}$ is the magnetic field strength, and Re$[\,]$ represents taking the real part. Without loss of generalization, the parameters for plotting are chosen to be  $z_0=25\lambda_0$, and $z=0.1\lambda_0$.

Now it has been proven that a simple air-nonreciprocal material planar surface can lead to asymmetric radiation distribution. In the following context, I focus specifically on the case presented in \textbf{Figure \ref{fig2}c}, where the material is nonreciprocal with $\epsilon_d=1.5+0.3\,i$ and $\epsilon_{xz}=1.5\,i$.

\subsection{Carrier-Envelope-Phase (CEP) manipulation}
\begin{figure}[h!]
\centering
  \includegraphics[width=0.5\linewidth]{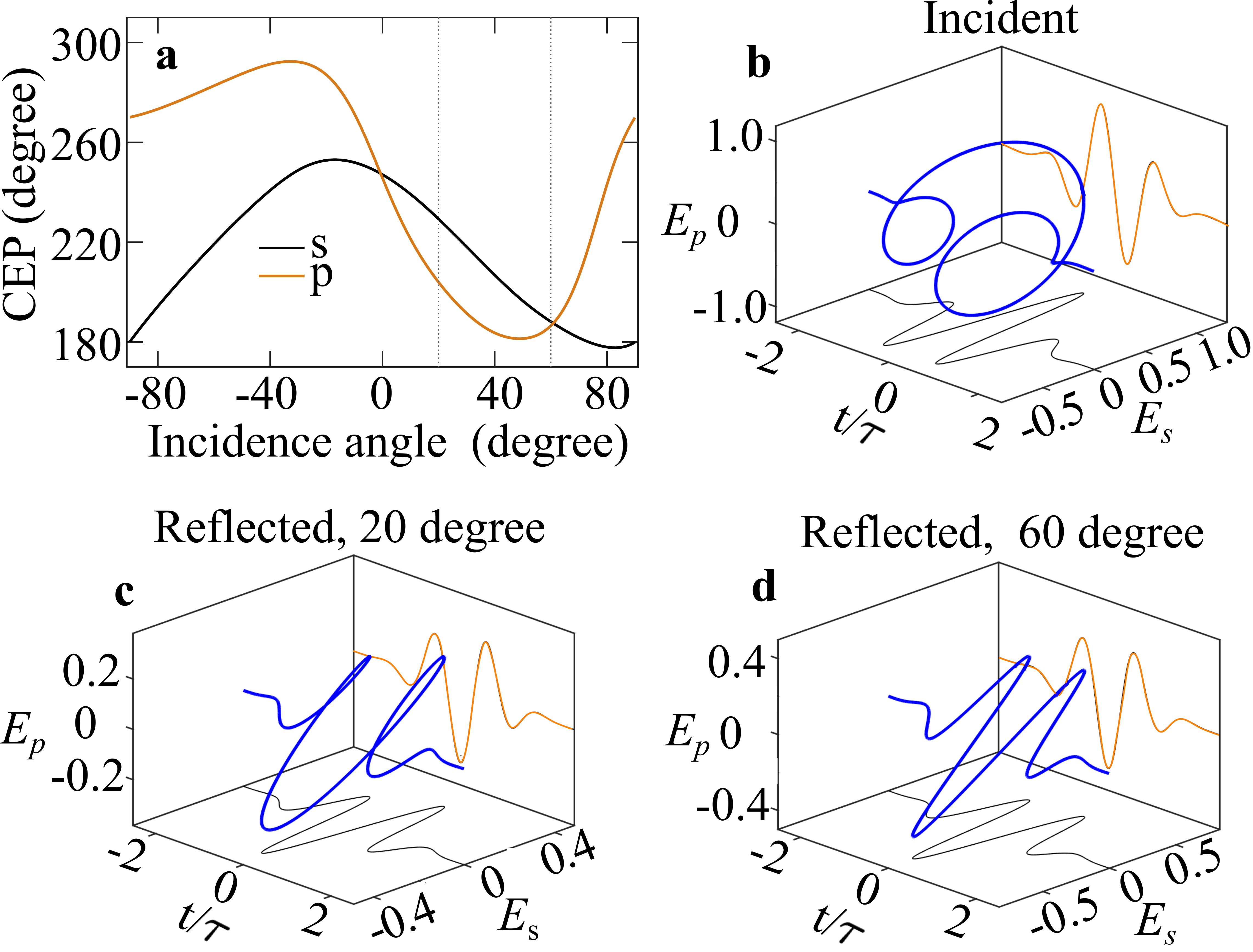}
\caption{Panel \textbf{a} displays the CEP of the s- and p- polarized light plotted against the incidence angle. The vertical dashed lines represent incidence angles 20 and 60 degrees, corresponding to the reflected electric field presented in \textbf{c} and \textbf{d}, respectively. Panel \textbf{b} shows the circularly polarized incidence field.}\label{fig:cep}
\end{figure}

Here we investigate how nonreciprocal material can be leveraged to manipulate the carrier-envelope-phase (CEP) of the pulse. Without loss of generalization, we look into one example with $\phi=45$ degree and the pulse duration related parameter $\tau=5/\omega_0$. By defining the amplitude of the s- and p- polarized light as
\begin{align}
 & E_s=\exp{(-t^2/\tau^2)}\exp{(iw_0t)},\\   & E_p=\exp{(-t^2/\tau^2)}\exp{(iw_0t+i\pi/2)}
\end{align}
one can write the input circularly polarized electric field as $\bm{E}_\text{in}(t)=\rm Re$$[E_s\hat{s}+E_p\hat{p}]$ (see \textbf{Figure \ref{fig:cep}b}), where $\hat{s}$ ($\hat{p}$) is the unit vector for s-polarisation (p-polarisation). Since $\bm{E}_\text{in}(t)$ is circularly polarized, the CEP of the $E_s$ and $E_p$ are 0 and $pi/2$ (90 degrees), respectively. The electric field after reflection from the nonreciprocal material surface is written as 
\begin{equation}
E_\text{out}(t)=\text{Re}[r_{sp}E_s\hat{p}+r_{ps}E_p\hat{
s}+r_{pp}E_p\hat{p}+r_{ss}E_s\hat{s}].
\end{equation}

\textbf{Figure \ref{fig:cep}a} shows how the CEP of s- and p- polarized light change as a function of incidence angle. In particular, two incidence angles 20 degrees (\textbf{Figure \ref{fig:cep}c}) and 60 degrees (\textbf{Figure \ref{fig:cep}d}) corresponding to elliptically and linearly polarized light after reflection, are presented as examples. These two angles are marked by the vertical dashed lines in Figure \ref{fig:cep}a.

The results suggest that the electric field transforms from a circularly polarized pulse into an elliptically polarized or linearly polarized pulse by simply reflecting from the nonreciprocal material surface. By simply changing the pulse incidence angle, continuous tuning of CEP is achieved. This brings alternatives for manipulating electromagnetic waveform, particularly in frequency ranges where traditional methods result in significant loss and beam distortions, such as the terahertz regime.

\subsection{Phase-Front Induced Spatial Profile Manipulation}
Here, I present how the nonreciprocal material can potentially be used to achieve spatial profile manipulation via phase-front manipulation. In this example, $\phi=0$ is chosen for simplicity, because in this configuration, $r_{sp}=r_{ps}=0$. With the p-polarized incident light, $r_{pp}$ is the only parameter of interest. The beam size related parameter is normalized to the wavelength ($\sigma=2\lambda_0$). The input electric field is at normal incidence and is defined as 
 \begin{equation}
     E_\text{in}(x)=\exp{[-(x/\sigma)^2]}\exp{(\psi)},
 \end{equation}
where the purely real part represents the spatial Gaussian envelope and the phase-front is noted by $\exp{(\psi)}$. Flat phase-front and quadratic phase-front are represented by $\psi=0$ and $\psi=\pm i(2x/\sigma) ^2$, respectively. 

\begin{figure}[h!]
\centering
  \includegraphics[width=0.4\linewidth]{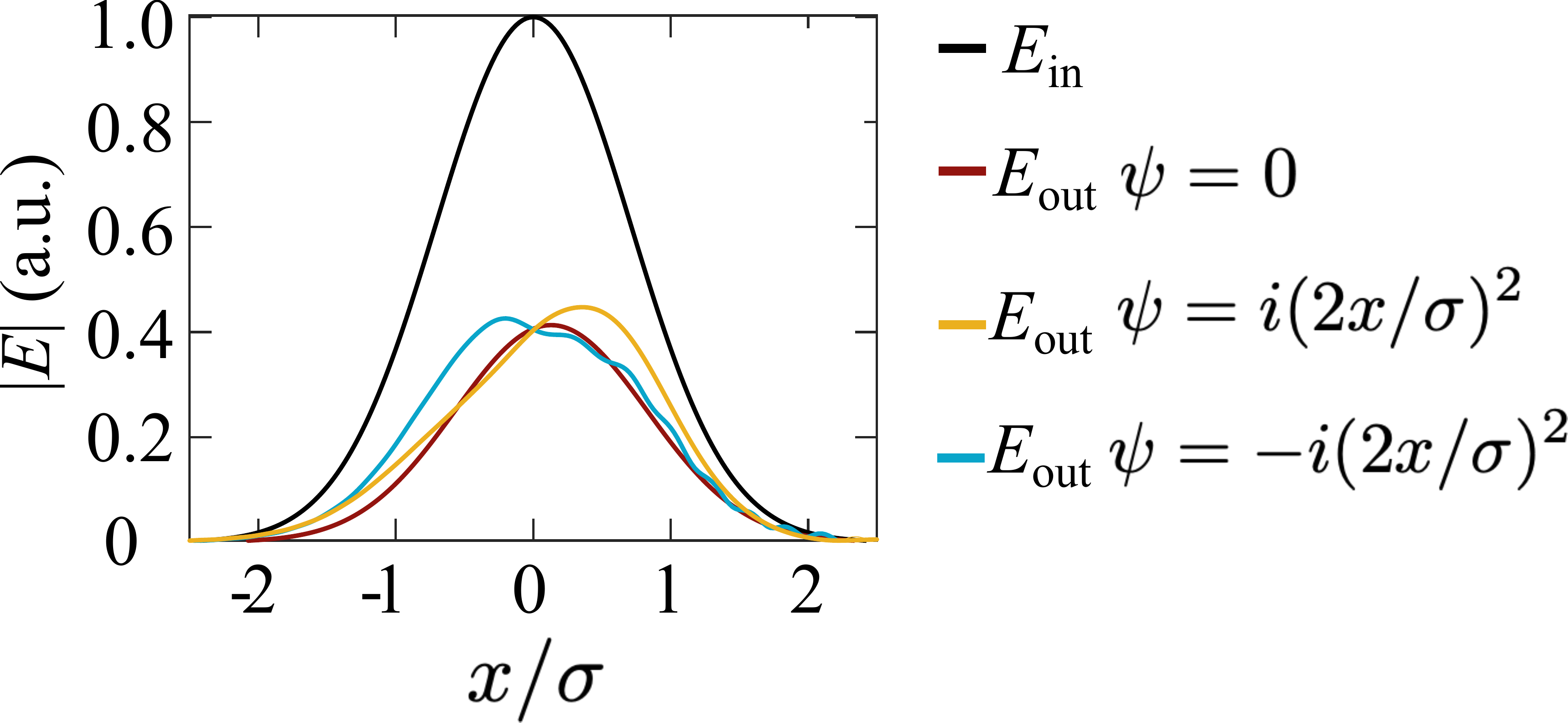}
\caption{Influence of the nonreciprocal material on the spatial profile of a Gaussian pulse upon reflection. The black curve represents the input pulse spatial profile. The red curve represents the reflected pulse without any additional phase. The yellow and blue curves represent the reflected pulse with quadratic phases of "+" and "-" signs, respectively.}\label{spatial}
\end{figure}

\textbf{Figure \ref{spatial}} shows that with a symmetric distribution of both the envelope and the phase-front, after reflection, an asymmetric spatial profile is obtained owing to the asymmetric response of the nonreciprocal material for $+\bm{k}$ and $-\bm{k}$. Since the space $x$ and momentum $k_x$ are a Fourier pair, owing to the spatial distribution in $x$, though the input field is at normal incidence, it still contains photons with different incidence angles i.e. a distribution in $k_x$. Remember that the reflection coefficient is an incidence angle dependent variable too, i.e. $r_{pp}(k_x)$. For more information on the calculations performed, please refer to Section S5 in the SM document. The results suggest that one can achieve spatial profile manipulation using phase-front manipulation.

\section{Conclusions}
These findings demonstrate that a simple planar interface between semi-infinite air and a nonreciprocal material enables effective spatio-temporal electromagnetic wave manipulation. The utilization of a single reflection on a planar surface presents a method with the potential for wide-frequency-range application. This characteristic is especially advantageous in frequency ranges like terahertz, where the avoidance of propagation through optical elements alleviates potential absorption, distortion, and diffraction. In addition, the closed-form expressions for a single dipole can be extended to the emission of arbitrary objects using the discrete dipole approximation.

These findings lay the foundation for pattern-free electric field manipulation using nonreciprocal materials, offering promising avenues for future advancements in this field.\\

\medskip
\noindent\textbf{Supporting Information} \par 
Supporting Information is available from the Wiley Online Library or from the author.\\

\medskip
\noindent \textbf{Acknowledgements:} \par 
Lu Wang thanks Prof. Xiaojun Liu for helpful discussions on this manuscript.

\medskip
\noindent\textbf{Conflict of Interest:} \par 
The author declares no conflict of interest.

\medskip
\noindent\textbf{Data Availability Statement:} \par 
The data and code are available upon request.

\bibliographystyle{MSP}
\bibliography{pic_bib/apssamp}

\end{document}

%% file: pic_bib/my_style.tex
\usepackage{amsmath} 
\usepackage{amssymb}  
\usepackage{amsfonts} 
\usepackage{dsfont}
\usepackage{bm}  
\usepackage{bbm}   
\usepackage{bbold}   
\usepackage{braket} 
\usepackage{color} 
\usepackage{comment}  
\usepackage{dcolumn}  
\usepackage{enumerate}  
\usepackage{epsfig}  
\usepackage{gensymb}  
\usepackage{graphicx}  \usepackage{indentfirst}  \usepackage{lmodern}  
\usepackage{mathrsfs}  
\usepackage{mathtools}  
\usepackage{psfrag}  
\usepackage{pst-all} 
\usepackage{soul}  
\usepackage{xcolor}
\usepackage{float} 
\usepackage[colorlinks,linkcolor=blue,citecolor=blue,urlcolor=blue,hyperindex,driverfallback=dvipdfm]{hyperref}  \usepackage[T1]{fontenc} 

\usepackage{dirtytalk}
\usepackage{CJKutf8}

 \usepackage{float} 
\usepackage{placeins}
 \usepackage{bm}
 \usepackage{multicols}

 \usepackage[normalem]{ulem} 
 \makeatletter
\newcommand*{\addFileDependency}[1]{
  \typeout{(#1)}
  \@addtofilelist{#1}
  \IfFileExists{#1}{}{\typeout{No file #1.}}
}
\makeatother

\usepackage[strict]{changepage}
\usepackage{calc}
\usepackage{dirtytalk}
\usepackage{pifont}

\usepackage{xr-hyper}

\usepackage{graphicx}
\usepackage{cancel}
\usepackage{longtable}
\usepackage{array}
\newcolumntype{C}[1]{>{\centering\arraybackslash}p{#1}}

\usepackage{ulem}
\usepackage{titlesec}
 \titleformat{\paragraph}[hang]{\bfseries}{}{0pt}{\uline}
\titlespacing*{\paragraph}
{0pt}{3.25ex plus 1ex minus .2ex}{1.5ex plus .2ex}

